\documentclass[letterpaper]{article} 
\usepackage{aaai24}  
\usepackage{times}  
\usepackage{helvet}  
\usepackage{courier}  
\usepackage[hyphens]{url}  
\usepackage{graphicx} 
\usepackage{natbib}  
\usepackage{caption} 
\usepackage{algorithm}
\usepackage{algorithmic}
\usepackage{bm}
\usepackage{enumitem}       
\usepackage{bbding}       
\usepackage{multirow}
\usepackage{amsmath}
\usepackage{amssymb}
\usepackage{newfloat}
\usepackage{listings}
\DeclareCaptionStyle{ruled}{labelfont=normalfont,labelsep=colon,strut=off} 
\floatstyle{ruled}
\newfloat{listing}{tb}{lst}{}
\floatname{listing}{Listing}
\title{Plug-in Diffusion Model for Sequential Recommendation}
\author{
    Haokai Ma\textsuperscript{\rm 1},
    Ruobing Xie \textsuperscript{\rm 3},
    Lei Meng \textsuperscript{\rm 2, \rm 1}\thanks{Corresponding Author.},
    Xin Chen \textsuperscript{\rm 3},
    Xu Zhang \textsuperscript{\rm 3},
    Leyu Lin \textsuperscript{\rm 3},
    Zhanhui Kang \textsuperscript{\rm 3}
}
\affiliations{
    \textsuperscript{\rm 1} School of Software, Shandong University, China \\
    \textsuperscript{\rm 2} Shandong Research Institute of Industrial Technology, China \\
    \textsuperscript{\rm 3} Tencent, China \\
    mahaokai@mail.sdu.edu.cn, ruobingxie@tencent.com, lmeng@sdu.edu.cn,\\ \{andrewxchen, xuonezhang, goshawklin, kegokang\}@tencent.com
}

\usepackage{bibentry}

\begin{document}

\maketitle

\begin{abstract}
Pioneering efforts have verified the effectiveness of the diffusion models in exploring the informative uncertainty for recommendation. Considering the difference between recommendation and image synthesis tasks, existing methods have undertaken tailored refinements to the diffusion and reverse process. However, these approaches typically use the highest-score item in corpus for user interest prediction, leading to the ignorance of the user's generalized preference contained within other items, thereby remaining constrained by the data sparsity issue. To address this issue, this paper presents a novel Plug-in Diffusion Model for Recommendation (PDRec) framework, which employs the diffusion model as a flexible plugin to jointly take full advantage of the diffusion-generating user preferences on all items. Specifically, PDRec first infers the users' dynamic preferences on all items via a time-interval diffusion model and proposes a Historical Behavior Reweighting (HBR) mechanism to identify the high-quality behaviors and suppress noisy behaviors. In addition to the observed items, PDRec proposes a Diffusion-based Positive Augmentation (DPA) strategy to leverage the top-ranked unobserved items as the potential positive samples, bringing in informative and diverse soft signals to alleviate data sparsity. To alleviate the false negative sampling issue, PDRec employs Noise-free Negative Sampling (NNS) to select stable negative samples for ensuring effective model optimization. Extensive experiments and analyses on four datasets have verified the superiority of the proposed PDRec over the state-of-the-art baselines and showcased the universality of PDRec as a flexible plugin for commonly-used sequential encoders in different recommendation scenarios. The code is available in https://github.com/hulkima/PDRec.
\end{abstract}

\section{Introduction}
Personalized recommendation aims to capture user preference from the massive user behaviors and predict the appropriate items the user will be interested in \cite{RS3,RS2,RS1}. Sequential recommendation (SR) is a effective method for inferring dynamic interests from the user's historical behavior sequences\cite{DGSR, MLP4Rec, ICLRec}. However, most users in the real world only interact with a limited number of items within the overall item corpus, consequently leading to the data sparsity problem \cite{xia2021knowledge, P2FCDR, chen2023enhanced}. 

\begin{figure}[!t]
\centering
\includegraphics[width=0.48\textwidth]{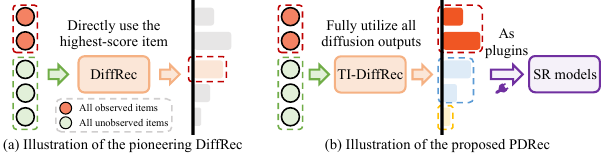}
\caption{Illustration of the difference between the pioneering DiffRec and PDRec, where each rectangle denotes the user's diffusion-based preference of the corresponding item.}
\label{fig:motivation}
\end{figure}

Diffusion model (DM), benefiting from its characteristics of diverse representation and informative uncertainty, has achieved state-of-the-art results in the field of image synthesis \cite{DDPM}, semantic segmentation \cite{SS_v1}, and time series imputation \cite{TSL_v1}. This demonstrates the dominance of DM as a novel generative paradigm in multiple generation tasks. Looking back to the real-world recommender systems, it could be regarded as a generator of the complete user-item interaction matrix based on the extremely sparse supervised signals \cite{CoMix}. It prompts an intuitive question: \textbf{\emph{Can we take full advantage of DM's potent generalization capability to generate user preferences on both observed and unobserved items, thereby addressing the sparsity issue in recommendation?}}

Recently, CODIGEM \cite{CODIGEM} and DiffRec \cite{DiffRec} have introduced DM into recommendation, which generates users' preferences based on their historical behaviors, yielding promising results. However, these pioneering studies still grapple with two challenges: 
(1) \textbf{\emph{How to fully utilize the generalized user preferences from DM?}} As shown in Fig.\ref{fig:motivation} (a), these methods merely utilize the highest-scored item as the final prediction in recommendation, overlooking users' preferences towards other items in corpus and struggling with data sparsity issue. However, these preferences encapsulate substantial informative and generalized knowledge during the inference process of DM. (2) \textbf{\emph{How to incorporate the diffusion-based knowledge to construct a universal framework that could smoothly cooperate with different SR models?}} These DM-based methods are primarily proposed for collaborative filtering (CF), failing to fully integrate the time-aware sequential behavioral information. This leads to huge gaps between them and real-world recommenders, limiting their practical feasibility and universality (as indicated in Table \ref{tab:main_results}, even T-DiffRec exhibits notable disparities from traditional SR models).

To address these issues, we propose a novel \textbf{Plug-in Diffusion Model for Recommendation (PDRec)} framework, which leverages the diffusion models as a flexible plugin and make full use of the diffusion-generated user preferences on all items. The coarse overview of PDRec is illustrated in Fig.\ref{fig:motivation} (b). Specifically, we first present a time-interval diffusion model on the basis of T-DiffRec to facilitate the more precise generation of dynamic user preferences for both observed and unobserved items. 
These diffusion-based preferences on all items (i.e., preferences for observed items, top-ranked unobserved items, and low-scored unobserved items) are utilized for jointly guiding the effective and stable optimization direction: 
(a) We devise a \emph{Historical Behavior Reweighting (HBR)} method for users' historical behaviors, which identifies the high-quality behaviors and reduces noisy interactions via the generated preferences given by previous diffusion models.
(b) We also propose a \emph{Diffusion-based Positive Augmentation (DPA)} approach to convert the unobserved items with top-ranked diffusion-based preferences to the potential positive labels via self-distillation, bringing in additional high-quality and diverse positive signals to alleviate data sparsity. 
(c) To alleviate the potential false negative sampling issue, we design a \emph{Noise-free Negative Sampling (NNS)} strategy, which selects safer negative samples from the low-scored unobserved items provided by diffusion in training.
The advantages of PDRec include: (1) HBR facilitates the discovery of more informative supervised signals from the global diffusion aspect, which could better guide model optimization. (2) DPA and NNS introduce additional knowledge on unobserved items that alleviates data sparsity issues. (3) PDRec is effective, universal, and easy-to-deploy, which could be conveniently applied to different datasets, base models, and recommendation tasks.

Extensive experiments on four real-world datasets with three base SR models demonstrate that our proposed PDRec achieves significant and consistent improvements across various datasets and tasks, including SR and cross-domain SR. Furthermore, we conduct comprehensive ablation studies and universality analyses to validate the effectiveness and universality of all components in PDRec.
The main contributions of this paper are summarized as follows:
\begin{itemize}[leftmargin=*, topsep=0.2pt,parsep=0pt]
  \item We propose a model-agnostic Plug-in Diffusion Model for Recommendation, which fully leverages the diffusion-based preferences on all items to improve base recommenders. To the best of our knowledge, we are the first to integrate the diffusion model as a plugin for different types of recommendation models and downstream tasks.
  \item The proposed HBR, DPA, and NNS are effective, model-agnostic, and easy-to-deploy plug-in strategies, which involve informative diffusion-generating preferences on all items to alleviate data sparsity.
  \item Our PDRec achieves significant and consistent improvements on different datasets, base SR models, and tasks. Its detachable components are well-received in practice.
\end{itemize}

\section{Related Work}
\textbf{Sequential recommendation.}
Sequential Recommendation (SR) is one of the representative methods for capturing users' dynamic temporal-aware preference evolution patterns by modeling the sequential dependencies of their historical behaviors, thereby recommending the next item that the user may be interested in \cite{MLP4Rec}. 
In recent years, Convolutional Neural Network (CNN) \cite{RCNN, Caser}, Recurrent Neural Network (RNN) \cite{NARM, GRU4Rec} and Transformer \cite{BERT4Rec, SASRec} are introduced into SR to capture users' preference dependencies. GRU4Rec \cite{GRU4Rec} employs the Gate Recurrent Unit (GRU) as the sequential encoder to learn users' long-term dependencies. SASRec \cite{SASRec}, one of the most widely used methods in SR, introduces Transformers for historical behavior interaction modeling. 
CL4SRec \cite{CL4SRec} is a strong SR models that proposes three sequence-based augmentations to construct contrastive learning (CL) tasks in SR. Nevertheless, existing methods primarily focus on modeling users' long- and short-term behaviors with various neural architectures, disregarding the potential impact of the time-interval-sensitive knowledge and the recommender's generalization capability across the entire corpus on the modeling of user preferences.

\noindent
\textbf{Diffusion models in recommendation.}
As a prominent deep generative method, Diffusion Models (DM) are inspired by non-equilibrium statistical physics and hasve demonstrated exceptional performance in Super Resolution \cite{SR_v1, SR_v2}, Semantic Segmentation \cite{SS_v1}, and Time Series Imputation \cite{TSL_v2}. Despite this, the relevant studies in the field of recommendation are marked by a notable scarcity. CODIGEM \cite{CODIGEM} leverages DM to generate robust collaborative signals and latent representations by modeling intricate and non-linear patterns. DiffRec \cite{DiffRec} reduces the added noises into the generative process to retain globally analogous yet personalized collaborative information in a denoising manner. DiffuRec \cite{DiffuRec} and DiffRec* \cite{DiffRec*} corrupts the item representations into the Gaussian distribution and reverses them based on the historical behaviors to employ uncertainty injection in item representation construction. 

However, the former two DM works exhibit an excessive reliance on the top-ranked data derived from diffusion, not only resulting in computing consumption and homogeneous recommendation results but also disregarding the comprehensive user historical behaviors. 
While the latter share the same modeling pipeline with SR methods, thus can serve as the base SR model within PDRec. The proposed PDRec differs from these works: (a) instead of directly training a diffusion model, we smartly leverage a pre-trained DM model to diminish the time complexity in model training. (b) We achieve the dual enhancement of recommendation diversity and preference modeling through denoising, knowledge distillation and negative sampling on sequence encoders with the diffusion-based preference. (c) PDRec is model- and task-agnostic, enabling its application across different sequence encoders and recommendation scenarios.

\section{Time-Interval Diffusion Model}
\label{sec.TI-DiffRec}
\begin{figure}[!t]
\centering
\includegraphics[width=0.45\textwidth]{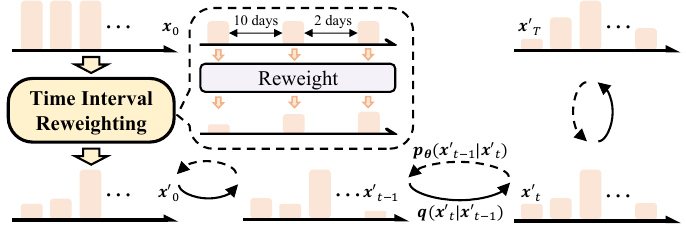}
\caption{The illustration of our enhanced time-interval diffusion recommendation model (TI-DiffRec).}
\label{fig:dm}
\end{figure}

In this section, we present the Time-Interval Diffusion Recommendation model (TI-DiffRec). Following the classical DM methods \cite{DDPM, IDDPM}, the pioneering methods \cite{CODIGEM, DiffRec} typically leverage the original interaction matrix for diffusion, which is challenging to apply in SR. Despite incorporating the temporal order of user interactions, T-DiffRec \cite{DiffRec} overlooks the time interval between consecutive behaviors, potentially leading to the issue of preference drift. As illustrated in Fig.\ref{fig:dm}, we introduce an additional process of time interval reweighting alongside the diffusion and reverse process to tackle this challenge.

\begin{itemize}[leftmargin=*, topsep=0.2pt,parsep=0pt]
  \item \textbf{Time Interval Reweighting.} To incorporate the time interval information into diffusion, we first generate the time interval-aware input $\bm{x}'_{0}$ with the behavior sequence $S_{u}\!=\!\{i^{1}_{u}, i^{2}_{u}, \cdots, i^{p}_{u}\}$ and the corresponding timestamps $T_{u}\!=\!\{t^{1}_{u}, t^{2}_{u}, \cdots, t^{p}_{u}\}$ of user $u$. Subsequently, we compute the time-interval weight $w^{j}_u=w_{\min}+\frac{t^{j}_u-t^{1}_u}{t^{p}_u-t^{1}_u}\left(w_{\max}-w_{\min}\right)$ of each behavior $i^{j}_{u}$, 
  where $w_{\min}$ and $w_{\max }$ denote the predefined lower and upper bounds. Thus we define $\bm{x}'_{0}\!=\![x_{1}, x_{2}, \cdots, x_{|I|}]$ as the initial state for diffusion, where $x_{i^{j}_{u}}\!=\!w^{k}_{u}$ or $0$ indicates whether $u$ has interacted with $i^{j}_{u}$ or not ($k$ denotes the index corresponding to $i^{j}_{u}$ within $S_{u}$).
  \item \textbf{Diffusion Process.} Generally, the diffusion process gradually injects uncertainty noise into the original data until the fully disordered state. The significant difference between DM and other latent variable models is that the transition kernel $q\left(\bm{x}'_t | \bm{x}'_{t-1}\right)$ used in DM obtains latent variables $\bm{x}'_t$ in a Markov chain process. Specifically, we employ the Gaussian perturbation as the transition kernel $q\left(\bm{x}'_t | \bm{x}'_{t-1}\right):=\mathcal{N}\left(\bm{x}'_t;\sqrt{1-\beta_t}\bm{x}'_{t-1}, \beta_t \mathbf{I}\right)$, where the variances $\beta_t\!\in\! (0,1)$ controls the Gaussian noise scales added at the step $t$. Note that the typical DM methods \cite{DDPM} fix the variances above to constants, demonstrating the notable property of DM that $q$ has no learnable parameters. Thus, we can directly generate $q\left(\bm{x}'_t\! \mid\! \bm{x}'_0\right)\!:=\!\mathcal{N}\left(\bm{x}'_t ; \sqrt{\bar{\alpha}_t} \bm{x}'_0,\left(1-\bar{\alpha}_t\right) \mathbf{I}\right)$ with the notation of $\alpha_t:=1-\beta_t, \bar{\alpha}_t:=\prod_{t^{\prime}=1}^t \alpha_{t^{\prime}}$ and $t\!\in\![1,2,\cdots,T]$. If $T\!\rightarrow\!+\infty$, $\bm{x}'_T$ asymptotically converges to the standard Gaussian distribution. That is, given $\bm{x}'_{0}$, we can easily obtain $\bm{x}'_{t}=\sqrt{\bar{\alpha}_t} \bm{x}'_0+\sqrt{1-\bar{\alpha}_t} \bm{\epsilon}$ by sampling the Gaussian vector $\bm{\epsilon} \sim \mathcal{N}(\mathbf{0}, \bm{I})$ via the reparameterization \cite{VAE}. 
  \item \textbf{Reverse Process.} The reverse process aims to recover user's interactions step by step from the standard Gaussian distribution through the denoising transition. Precisely, given the Gaussian distribution vector $\bm{x}'_T$, we gradually remove the noise and recover the original interactions with the learnable transition kernel $p_\theta\left(\bm{x}'_{t-1}\!\mid\!\bm{x}'_t\right)$ in the reverse direction. The reverse transition phase can be defined as:
  \begin{equation}
      p_\theta\left(\bm{x}'_{t-1} \mid \bm{x}'_t\right):=\mathcal{N}\left(\bm{x}'_{t-1} ; \boldsymbol{\mu}_\theta\left(\bm{x}'_t, t\right), \boldsymbol{\Sigma}_\theta\left(\bm{x}'_t, t\right)\right)
  \end{equation}
  where $\!\boldsymbol{\mu}_\theta\left(\bm{x}'_t, t\right)\!$ and $\!\boldsymbol{\Sigma}_\theta\left(\bm{x}'_t, t\right)\!$ are parameterized by Deep Neural Networks (DNN) and  $\theta$ denotes model parameters. We can model complex interaction generation procedures for recommendation by such an iterative reverse process.  
\end{itemize}

\section{Plug-in Diffusion Model}

\subsection{Task Formulation and Overall Framework}
SR aims to improve the next-item recommendation performance via the users’ historical sequential behaviors. To this end, given the \emph{behavior sequence} $S_{u}=\{i^{1}_{u}, i^{2}_{u}, \cdots, i^{p}_{u}\}$ of user $u\!\in\!\mathcal{U}$, where $i_j\!\in\!\mathcal{I}$ is the $j$-th behavior of $u$ and $p$ denote the historical behavior length, PDRec tries to recommend the target item $i^{p+1}_u$ that will be interacted by this user.

\begin{figure}[!t]
\centering
\includegraphics[width=0.48\textwidth]{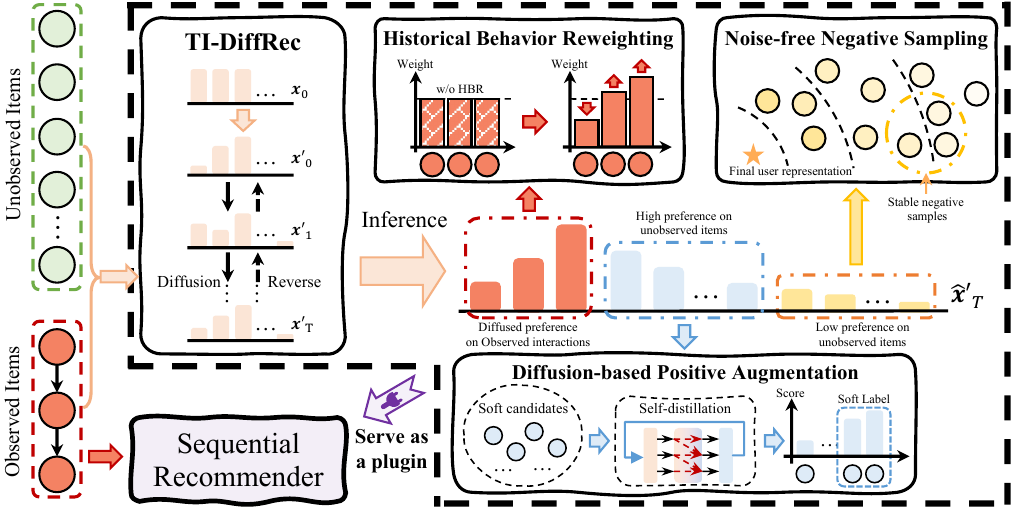}
\caption{The overall structure of the proposed PDRec.}
\label{fig:overall_structure}
\end{figure}

In this section, we describe the proposed model-agnostic Plug-in Diffusion Model for Recommendation (PDRec) framework, which leverages the diffusion model as a flexible plugin to accurately model the dynamic user preferences in SR. The overall structure of PDRec is illustrated in Fig.\ref{fig:overall_structure}. Specifically, PDRec first explicitly regards the user behavior timestamps in diffusion to capture the dynamics of the actual sequential patterns and generate the user's preferences on all items. Next, PDRec proposes a Historical Behavior Reweighting (HBR) strategy to identify specific indispensable supervised signals with the diffusion-based preferences on observed items. Additionally, PDRec designs a Diffusion-based Positive Augmentation (DPA) method to allievate the data saprsity problem, which conducts self-distillation to incorporate the probable interactions from the unobserved items as the augmented soft samples into the training process in a dynamic manner. Finally, PDRec employs Noise-free Negative Sampling (NNS) to select stable negative samples, with the aim to mitigate the potential false negative problem. It is noteworthy that PDRec is task-agnostic, allowing such a framework can be easily migrated to cross-domain sequential recommendation (CDSR) tasks, and the related analysis is illustrated in Sec.\ref{sec.uni_scen}.

\subsection{Historical Behavior Reweighting}
The basic task of sequential preference modeling is to accurately and comprehensively leverage the historical behavior sequences. It is intuitive that different items within the user's sequential sequence should hold varying degrees of importance for the next-item prediction. Therefore, the critical aspect of historical preference modeling lies in the adaptive and fine-grained differentiation of all observed items. This involves integrating the distinct importance of various items into the training phase. To this end, we propose a Historical Behavior Reweighting (HBR) strategy to reweight the supervised signals in training for behavior sequence denoising.

Specifically, given the pre-trained TI-DiffRec model and the complete interaction state $\bm{x}_{0}$ of user $u$ in the training set, we first obtain the time-interval-aware state $\bm{x'}_{0}$.
Following the inference process in DiffRec \cite{DiffRec}, we regard the $\bm{x'}_{0}$ which is naturally noisy and retaining personalized information as the noised state $\bm{\hat{x}'}_{T}$.
Then we employ the reverse denoising by $\bm{\hat{x}'}_{T}\!\rightarrow\!\cdots\!\rightarrow\!\bm{\hat{x}'}_{1}\!\rightarrow\!\bm{\hat{x}'}_{0}$ to generate the diffusion-based preferences $\bm{\hat{x}'}_{0}\!\in\!\mathbb{R}^{|\mathcal{I}|}$ of $u$ on all items $\mathcal{I}$. To reweight the supervised signals, we exclusively focus on the observed preferences $\bm{o}_{u}\!\in\!\mathbb{R}^{|\mathcal{I}^{+}_u|}$ and their corresponding ranking $\bm{r}_{u}\!\in\!\mathbb{R}^{|\mathcal{I}^{+}_u|}$ from $\bm{\hat{x}'}_{0}$ in HBR. The reweight vector $\bm{w}_u$ to the supervised signals of user $u$ is formulated as:
\begin{equation}
    \bm{\hat{w}}_u\!=\!(1\!-\!\omega_r)\cdot\omega_s\cdot\frac{\bm{o}_{u}\!-\!\min_{\bm{o}_{u}}}{\max_{\bm{o}_{u}}\!-\!\min_{\bm{o}_{u}}}+\omega_r\!\cdot\!\frac{1\!+\!\max_{\bm{r}_{u}}\!-\!\bm{r}_{u}}{\max_{\bm{r}_{u}}}
\end{equation}
where $\omega_s\!=\!len(S_{u})\!/\!sum(\frac{\bm{o}_{u}\!-\!\min_{\bm{o}_{u}}}{\max_{\bm{o}_{u}}\!-\!\min_{\bm{o}_{u}}})$, $\omega_r$ and $(1\!-\!\omega_r)$ denote the ranking weight of observed ranking and preferences respectively. 
This mixture ensures that the recommender's optimization direction remains reasonably aligned with the observed prior to the reweighting process. Finally, PDRec generates the final reweight vector $\bm{w}_u = \omega_f\cdot\min(\max(c_w, \min(\bm{\hat{w}}_u)), \max(\bm{\hat{w}}_u))$ by truncating and rescaling the reweight vector via the truncate value $c_w$ and the rescale weight $\omega_f$ to prevent certain signals dominate the optimization process. Note that PDRec only employs the inference process before model training without introducing excessive computational cost. With HBR, we can not only directly focus on the time-interval-aware preferences related to the user’s behaviors, but also leverage DM's informative uncertainty to denoise the dispensable items and highlight the indispensable actions in the user's behavior sequence.

\subsection{Diffusion-based Positive Augmentation}
Inspired by the promising performance of TI-DiffRec in SR and its inherent strong generalization characteristics, PDRec assumes that the user's diffusion-based preferences $\bm{\hat{x}'}_{0}$ on unobserved items encompass specific samples that user $u$ is potentially interested in but have never seen before. This intuitive observation exhibits heightened prominence within the top-ranked range of the user's diffusion-based preference. To transfer the generalized knowledge from the pre-trained TI-DiffRec to the sequential recommender, PDRec intelligently designs a Diffusion-based Positive Augmentation (DPA) method to distil the essential diffusion-based information by regarding the unobserved items with high preferences as the potential positive samples during training. 

To do it, PDRec first takes top-ranked $m$ items to form the potential soft samples $t_{u}$ based on the diffusion-based unobserved preferences $\bm{u}_{u}\!=\!\bm{\hat{x}'}_{0}\!\backslash\!\bm{o}_{u}$, where $\bm{\hat{x}'}_{0}$ and $\bm{o}_{u}$ denote the diffusion-based preferences of the corpus and supervised signals.
Following the assumption that "the last behavior in a user's behavioral sequence reflects his/her overall interests", PDRec calculate the matching score 
$m_u=[(\bm{h}_u)^\top\bm{t}_1,(\bm{h}_u)^\top\bm{t}_2,\cdots,(\bm{h}_u)^\top\bm{t}_m]$
between the user's last behavior representation $\bm{h}_u$ obtained by SASRec \cite{SASRec} and the item embedding matrix $\bm{T}_u=[\bm{t}_1,\bm{t}_2,\cdots,\bm{t}_m]$ of the potential soft samples $t_{u}$. After re-ranking the matching score $m_u$, PDRec extract the top-ranked $n$ items to obtain the soft positive augmentations $s_{u}$. The optimization approach for $s_{u}$ will be defined in Sec. \ref{sec.opt}.

\subsection{Noise-free Negative Sampling}
Existing recommendation algorithms generally require both positive and negative examples to model users' personalized preferences. They expect explicit interactions from the dataset ideally. However, explicit feedback is not always available in real-world scenarios, the ubiquitous users' implicit interactions may not necessarily reflect their real interests. Conventional recommenders typically employ uniform probability for negative sampling, which fail to consider the dynamic shifts in user preferences, potentially leading to the false negative problem to some extent. Inspired by the exploration of negative sampling strategies in recommendation \cite{DNS+, RealHNS}, PDRec introduces the Noise-free Negative Sampling (NNS) strategy to prioritize the unobserved samples with low-scored diffusion-based preference and select safe negative samples to direct HBR and DPA in the stable optimization direction.

In contrast to the DPA utilizing the unobserved items with high preferences as the soft positive augmentations, NNS creatively regard these items with low preferences as the additional negative samples in training. Precisely, given the diffusion-based unobserved preferences $\bm{u}_{u}$, PDRec sorts them, selects low-scored items from the unobserved corpus $\mathcal{I}^{-}_u=\mathcal{I}\backslash\mathcal{I}^{+}_u$, and assigns higher sampling probabilities to these items. The sampling probability of NNS is defined as:
\begin{equation}
 P^{N\!N\!S}(j\!\mid\!\mathcal{I}^{-}_u)\!=\!\left\{
    \begin{array}{rcl}
        \frac{1}{(1-\omega_m)l_u}, \!&&\! j \in K_u[\omega_m l_u:l_u]\\
        0,           \!&&\! j \in others\\
    \end{array} \right. 
\end{equation}
where $K_u$ is the re-ranked item list of $\mathcal{I}^{-}_u$ obtained by sorting the diffused unobserved preferences $\bm{u}_{u}$, $l_u\!=\!|K_u|$ denotes the number of unobserved items and $\omega_m$ denotes the initial proportion of the negative sampling. Note that the bigger the $\omega_m$ is, the more stable the samples will be drawn.

\subsection{Optimization Objectives}
\label{sec.opt}

\begin{table*}[!t]
\centering
\renewcommand\arraystretch{1.1}
\tabcolsep=1.0mm
\small
\begin{tabular}{|c|c|cc|ccc|ccc|ccc|}
\hline
\textbf{Datasets}    & \textbf{Metrics} & \textbf{T-DiffRec} & \textbf{TI-DiffRec} & \textbf{GRU4Rec} & \textbf{+PDRec} & \textbf{Improv.} & \textbf{SASRec} & \textbf{+PDRec} & \textbf{Improv.} & \textbf{CL4SRec} & \textbf{+PDRec} & \textbf{Improv.} \\ \hline
\multirow{6}{*}{\textbf{Toy}} 
    & N@1   & 0.1033 & 0.1058  & 0.0878 & \textbf{0.0899} & 2.39\% & 0.1095  & \textbf{0.1247} & 13.88\% & 0.1125  & \textbf{0.1254} & 11.47\% \\
    & N@5   & 0.1564 & 0.1618  & 0.1515 & \textbf{0.1617} & 6.73\% & 0.1779  & \textbf{0.2023} & 13.72\% & 0.1802 & \textbf{0.2041} & 13.26\% \\
    & N@10  & 0.1758 & 0.1823  & 0.1755 & \textbf{0.1879} & 7.07\% & 0.2020  & \textbf{0.2286} & 13.17\% & 0.2046 & \textbf{0.2305} & 12.66\% \\
    & HR@5  & 0.2055 & 0.2151  & 0.2128 & \textbf{0.2300} & 8.08\% & 0.2423  & \textbf{0.2752} & 13.58\% & 0.2438 & \textbf{0.2776} & 13.86\% \\
    & HR@10 & 0.2657 & 0.2787  & 0.2874 & \textbf{0.3112} & 8.28\% & 0.3169  & \textbf{0.3568} & 12.59\% & 0.3195 & \textbf{0.3595} & 12.52\% \\
    & AUC   & 0.5911 & 0.5968  & 0.5670 & \textbf{0.5909} & 4.22\% & 0.5771  & \textbf{0.6060} & 5.01\%  & 0.5805 & \textbf{0.6068} & 4.53\%  \\ \hline
\multirow{6}{*}{\textbf{Game}}  
    & N@1   & 0.1611 & 0.1746 & 0.1667 & \textbf{0.1808} & 8.46\% & 0.2111 & \textbf{0.2191} & 3.79\%  & 0.2106 & \textbf{0.2180} & 3.51\%  \\
    & N@5   & 0.2567 & 0.2723 & 0.2818 & \textbf{0.2996} & 6.32\% & 0.3310 & \textbf{0.3382} & 2.18\%  & 0.3294 & \textbf{0.3368} & 2.25\%  \\
    & N@10  & 0.2895 & 0.3040 & 0.3199 & \textbf{0.3380} & 5.66\% & 0.3682 & \textbf{0.3753} & 1.93\%  & 0.3682 & \textbf{0.3750} & 1.85\%  \\
    & HR@5  & 0.3451 & 0.3618 & 0.3893 & \textbf{0.4091} & 5.09\% & 0.4409 & \textbf{0.4475} & 1.50\%  & 0.4385 & \textbf{0.4456} & 1.62\%  \\
    & HR@10 & 0.4469 & 0.4600 & 0.5071 & \textbf{0.5282} & 4.16\% & 0.5559 & \textbf{0.5626} & 1.21\%  & 0.5584 & \textbf{0.5638} & 0.97\%  \\
    & AUC   & 0.7217 & 0.7234 & 0.7601 & \textbf{0.7786} & 2.43\% & 0.7865 & \textbf{0.7908} & 0.61\%  & 0.7857 & \textbf{0.7905} & 0.61\%  \\ \hline
\multirow{6}{*}{\textbf{Book}}  
    & N@1   & 0.3194 & 0.3275 & 0.3072 & \textbf{0.3359} & 9.34\% & 0.3594 & \textbf{0.3656} & 1.73\%  & 0.3554 & \textbf{0.3621} & 1.89\%  \\
    & N@5   & 0.4398 & 0.4491 & 0.4433 & \textbf{0.4757} & 7.31\% & 0.4948 & \textbf{0.5063} & 2.32\%  & 0.4942 & \textbf{0.5047} & 2.12\%  \\
    & N@10  & 0.4671 & 0.4776 & 0.4765 & \textbf{0.5091} & 6.84\% & 0.5272 & \textbf{0.5393} & 2.30\%  & 0.5276 & \textbf{0.5376} & 1.90\%  \\
    & HR@5  & 0.5459 & 0.5557 & 0.5643 & \textbf{0.6004} & 6.40\% & 0.6148 & \textbf{0.6306} & 2.57\%  & 0.6166 & \textbf{0.6304} & 2.24\%  \\
    & HR@10 & 0.6300 & 0.6435 & 0.6667 & \textbf{0.7033} & 5.49\% & 0.7150 & \textbf{0.7323} & 2.42\%  & 0.7197 & \textbf{0.7317} & 1.67\%  \\
    & AUC   & 0.8160 & 0.8202 & 0.8541 & \textbf{0.8728} & 2.19\% & 0.8790 & \textbf{0.8898} & 1.23\%  & 0.8820 & \textbf{0.8895} & 0.85\%  \\ \hline
\multirow{6}{*}{\textbf{Music}} 
    & N@1   & 0.3401 & 0.3494 & 0.3299 & \textbf{0.3540} & 7.31\% & 0.3753 & \textbf{0.3826} & 1.95\%  & 0.3689 & \textbf{0.3755} & 1.79\%  \\
    & N@5   & 0.4709 & 0.4773 & 0.4725 & \textbf{0.5000} & 5.82\% & 0.5170  & \textbf{0.5283} & 2.19\%  & 0.5096 & \textbf{0.5211} & 2.26\%  \\
    & N@10  & 0.4987 & 0.5049 & 0.5069 & \textbf{0.5348} & 5.50\% & 0.5503 & \textbf{0.5620} & 2.13\% & 0.5435 & \textbf{0.5558} & 2.26\%  \\
    & HR@5  & 0.5852 & 0.5886 & 0.5987 & \textbf{0.6287} & 5.01\% & 0.6421 & \textbf{0.6573} & 2.37\%  & 0.6353 & \textbf{0.6504} & 2.38\%  \\
    & HR@10 & 0.6706 & 0.6738 & 0.7048 & \textbf{0.7361} & 4.44\% & 0.7447 & \textbf{0.7612} & 2.22\%  & 0.7400 & \textbf{0.7573} & 2.34\%  \\
    & AUC   & 0.8329 & 0.8318 & 0.8768 & \textbf{0.8908} & 1.60\% & 0.8962 & \textbf{0.9040}  & 0.87\% & 0.8939 & \textbf{0.9026} & 0.97\%  \\ \hline
\end{tabular}
\caption{Results between backbones and PDRec on four datasets. All improvements are significant ($p \textless 0.05$ with paired t-tests).}
\label{tab:main_results}
\end{table*}

\begin{table}[t]
\centering
\small
\renewcommand\arraystretch{1.05}
\tabcolsep=1.6mm
\begin{tabular}{|c|cc|cc|}
\hline
\textbf{Datasets} & \multicolumn{1}{c|}{\textbf{Toy}} & \multicolumn{1}{c|}{\textbf{Game}} & \multicolumn{1}{c|}{\textbf{Book}} & \multicolumn{1}{c|}{\textbf{Music}}     \\ \hline
\begin{tabular}[c]{@{}c@{}} \textbf{Users}\end{tabular} &
  \multicolumn{1}{c|}{7,996} &
  \multicolumn{1}{c|}{7,996} &
  \multicolumn{1}{c|}{12,170} &
  \multicolumn{1}{c|}{12,170} \\ \hline
\textbf{Items} & \multicolumn{1}{c|}{37,868}  & \multicolumn{1}{c|}{11,735}   & \multicolumn{1}{c|}{33,697}   & \multicolumn{1}{c|}{30,707}   \\ \hline
\textbf{Records} & \multicolumn{1}{c|}{114,487}  & \multicolumn{1}{c|}{82,871} & \multicolumn{1}{c|}{514,015}  & \multicolumn{1}{c|}{558,352}  \\ \hline
\textbf{Density} & \multicolumn{1}{c|}{0.0378\%} & \multicolumn{1}{c|}{0.0883\%} & \multicolumn{1}{c|}{0.1253\%} & \multicolumn{1}{c|}{0.1494\%} \\ \hline
\end{tabular}
\caption{Statistics of four SR datasets.}
\label{tab:Dataset}
\end{table}

We calculate the predicted probability $\hat{y}\!=\!(\bm{h}_u)^\top \bm{v}_{q+1}$ with the sequence representation $\bm{h}_u$ of user $u$ and the item embedding $\bm{v}_{q+1}$. Then we formulate the Binary Cross-Entropy loss $\mathcal{L_{R}}$ and the self-distillation loss $\mathcal{L}_{D}$ in DPA as follows:
\begin{gather}
    \mathcal{L}_{R}\!=\!-\!\sum_{(u,i)\in \!R}\!\left[\bm{w}_u\!\cdot\!y_{u,i}\!\log \hat{y}_{u,i}\!+\!\left(1\!-\!y_{u,i}\right)\!\log\!\left(\!1\!-\!\hat{y}_{u,i}\!\right)\!\right]\\
    \mathcal{L}_{D}\!=\!-\!\sum_{(u,i) \in R^{+}} \left[y_{u,i}\!\log\!\hat{y}_{u, i}\right]
    \label{eq.L_CTR}
\end{gather}
where $R$ denotes the training set which contains the supervised signals, the random negative samples and the safe negative items sampled within NNS, $R^{+}$ denotes the soft positive augmentations $\!s_{u}\!$ in DPA, $\!\bm{w}_u\!$ is the final reweight vector in HBR, $y_{u,i}\!=\!1\slash0$ denote the positive and the sampled negative pairs respectively, and $\!\hat{y}_{u,i}\!$ denotes the predicted probability of $\!(u,i)$. To optimize in conjunction with the self-distillation augmentation, the objective function $\mathcal{L}$ is a linear combination of $\mathcal{L}_{R}$ and $\mathcal{L}_{D}$ with the loss weight $\omega_d$ of $\mathcal{L}_{D}$:
\begin{equation}
    \mathcal{L} = \mathcal{L}_{R} + \omega_d \mathcal{L}_{D}.
\end{equation}

\section{Experiments}
In this section, we conduct extensive experiments and analyses for the following research questions: 
(RQ1): How does PDRec perform against the state-of-the-art SR baselines (see Sec.\ref{sec.performance})? 
(RQ2): How do different components of PDRec benefit its performance (see Sec.\ref{sec.ablation})?
(RQ3): Is PDRec still effective with other base SR models (see Sec.\ref{sec.uni_seq})?
(RQ4): Could PDRec be further adopted to other tasks such as cross-domain sequential recommendation? (see Sec.\ref{sec.uni_scen})?

\subsection{Experimental Settings}
\subsubsection{Dataset.}
We conduct extensive experiments on four real-world datasets. We select ``Toys and Games'' and ``Video Games'' to form the \emph{Toy} and \emph{Game} dataset from Amazon \cite{amazon}. From Douban, we pick ``Books'' and ``Musics'' to form the \emph{Book} and \emph{Music} dataset \cite{douban}. 


\subsubsection{Baselines.}
We implement PDRec on three representative SR models: \textbf{GRU4Rec} \cite{GRU4Rec}, \textbf{SASRec} \cite{SASRec} and \textbf{CL4SRec} \cite{CL4SRec}, and compare it with \textbf{T-DiffRec} \cite{DiffRec} to validate its effectiveness and universality. Note that T-DiffRec \cite{DiffRec} is one of the SOTA DM-based recommenders that captures the temporal patterns in user interactions.

\begin{figure*}[!t]
\centering
\includegraphics[width=0.98\textwidth]{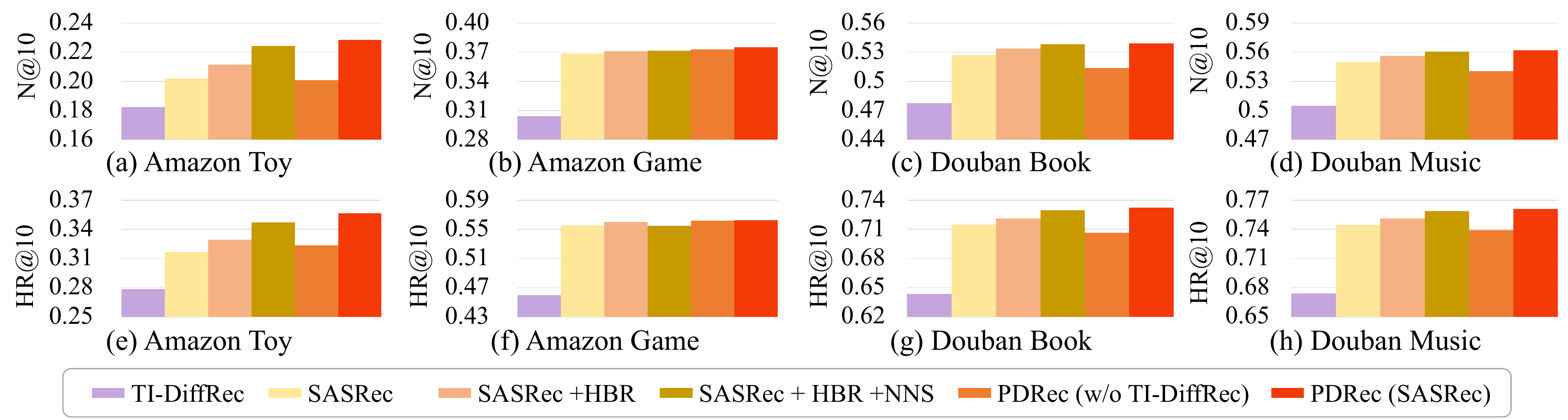}
\caption{Results on ablation study of PDRec (SASRec) on four datasets. Generally, all components are effective.}
\label{fig:ablation}
\end{figure*}

\subsubsection{Parameter settings.}
For fair comparisons, we set the learning rate and the maximum sequence length as 5e$^{-3}$ and 200. According to the natural distribution of behaviors, we set the $\omega_m$ as 0.5 for the relatively sparse Amazon datasets and 0.8 for the denser Douban datasets. Similarly, we define the number of coarse-grained sorted items $m$, the number of fine-grained resorted items $n$, and the loss weight $\omega_d$ of $\mathcal{L}_{SD}$ as $50$, $5$ and $0.3$ for Amazon. For Douban, these parameter are configured as $100$, $1$, and $0.01$, respectively. Due to the variations in TI-DiffRec's confidence range, PDRec exhibits minor discrepancies in the parameters of HBR across diverse datasets. That is, the ranking weight $\omega_r$, the truncate value $c_w$ and the rescale weight $\omega_f$ are denoted as $0.1$, $3$ and $2$ for Toy, $0.1$, $5$ and $4$ for Game, $0.3$, $3$ and $4$ for Book and $0.1$, $5$ and $2$ for Music. Each experiment is conducted five times with random seeds, and we report the average results.

\subsection{Performance Comparison on SR (RQ1)}
\label{sec.performance}
We conduct the experiments on four public datasets, adopting three typical evaluation metrics, including NDCG@k (N@k), Hit Rate@k (HR@k), and AUC with different $k=1,5,10$. Following \cite{SASRec}, we randomly sample 99 negative items for each positive instance in testing. Table \ref{tab:main_results} shows the overall performance comparison results, the best results of the same backbone are in boldface. It reveals the following observations:

(1) In general, PDRec significantly outperforms all baselines on four datasets, exhibiting the significance level $p \textless 0.05$ and the average error range $\leq 0.004$. This indirectly confirms the significance of (a) the observed interactions denoising is able to guide the recommender toward an accurate and unbiased optimization direction; (b) the handling of positive and negative aspects of unobserved interactions effectively leverages the informative yet user-imperceptible knowledge from the diffusion model, expanding user interests while stabilizing the training process. 

(2) Comparing the improvements across various datasets, we discover that PDRec benefits the relatively sparse Toy and Game datasets more. Meanwhile, PDRec can obtain promising performance even on denser datasets. Furthermore, we also observed that PDRec, implemented with diverse backbones, consistently exhibits significant improvements over its respective backbones. This may be attributed to the precise utilization of diffusion models, PDRec can assist in highlighting the actual long- and short-term sequential dependencies. As a task-agnostic framework, we further expand PDRec into the field of CDSR to employ the feasibility analyses on recommendation scenarios in Sec. \ref{sec.uni_scen}.

(3) Simultaneously, we notice the significant improvement of the proposed TI-DiffRec relative to T-DiffRec \cite{DiffRec}, underscoring the necessity of time-interval knowledge in SR. Nevertheless, the performance of these DM-based algorithms remains inferior to the existing SOTA sequential recommendation algorithms. In conjunction with the notable improvement of PDRec over these SR methods, the effectiveness of the proposed PDRec is firmly established. It can smartly combine the sequential modeling capability of (the future advanced) SR models and the potent generalized ability of diffusion models on the corpus, thus precisely accomplishing sequential recommendation tasks.

\begin{figure*}[!t]
\centering
\includegraphics[width=0.9\textwidth]{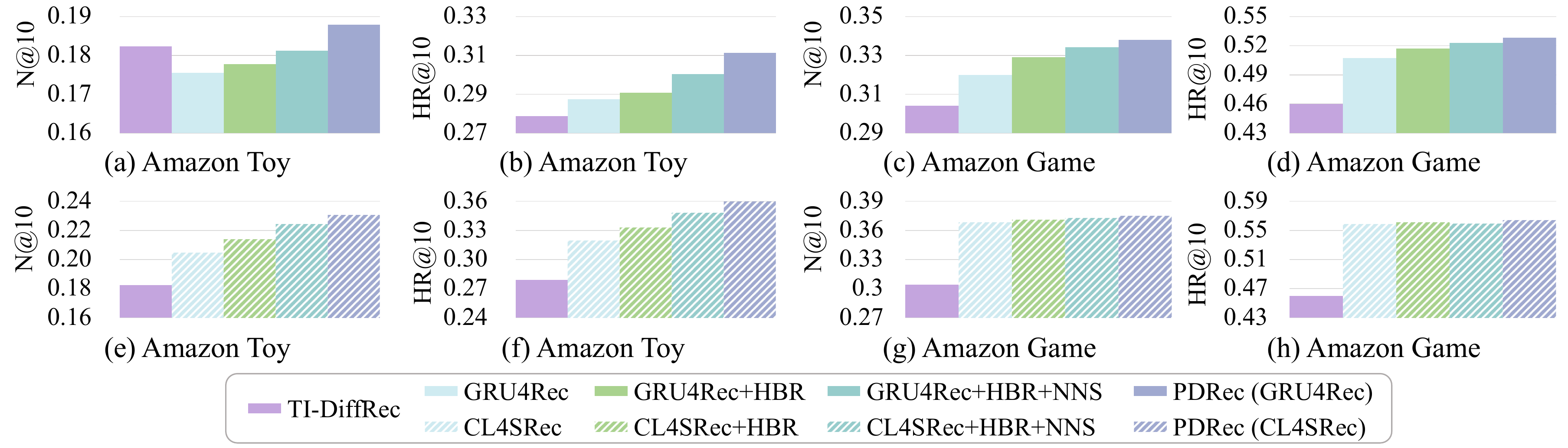}
\caption{Results of PDRec on GRU4Rec/CL4SRec and their ablation versions on Toy and Game datasets.}
\label{fig:uni_seq}
\end{figure*}

\begin{table*}[!t]
\centering
\small
\renewcommand\arraystretch{1.1}
\tabcolsep=1.6mm
\begin{tabular}{|c|c|c|c|c|c|c|c|c|c|c|c|}
\hline
\textbf{Setting} &
  \textbf{Algorithms} &
  \textbf{N@1} &
  \textbf{N@5} &
  \textbf{N@10} &
  \textbf{N@20} &
  \textbf{N@50} &
  \textbf{HR@5} &
  \textbf{HR@10} &
  \textbf{HR@20} &
  \textbf{HR@50} &
  \textbf{AUC} \\ \hline
\multirow{6}{*}{\begin{tabular}[c]{@{}c@{}}Game\\ ↓\\ Toy\end{tabular}}  
    & \textbf{T-DiffRec (M)}  & 0.0981 & 0.1520  & 0.1727 & 0.1934 & 0.2375 & 0.2029 & 0.2673 & 0.3494 & 0.5780  & 0.5924 \\ 
    & \textbf{TI-DiffRec (M)} & 0.1053 & 0.1598 & 0.1806 & 0.2008 & 0.2407 & 0.2111 & 0.2759 & 0.3562 & 0.5623 & 0.5932 \\ \cline{2-12} 
    & \textbf{SASRec (M)} & 0.1267 & 0.2019 & 0.2261 & 0.2490  & 0.2785  & 0.2722 & 0.3472 & 0.4380  & 0.5873 & 0.5951 \\ 
    & \textbf{+HBR}        & 0.1283 & 0.2061 & 0.2311 & 0.2533 & 0.2835  & 0.2785 & 0.3558 & 0.4438 & 0.5972 & 0.6092 \\ 
    & \textbf{+HBR+NNS}     & \underline{0.1264} & \underline{0.2068} & \underline{0.2323} & \underline{0.2542} & \underline{0.2844} & \underline{0.2815} & \underline{0.3606} & \underline{0.4480} & \underline{0.6013} & \textbf{0.6123} \\ 
    & \textbf{+HBR+NNS+DPA}  & \textbf{0.1302} & \textbf{0.2093} & \textbf{0.2348} & \textbf{0.2574} & \textbf{0.2873} & \textbf{0.2826} & \textbf{0.3616} & \textbf{0.4515} & \textbf{0.6026} & \underline{0.6106} \\ \hline
\multirow{6}{*}{\begin{tabular}[c]{@{}c@{}}Toy\\ ↓\\ Game\end{tabular}} 
    & \textbf{T-DiffRec (M)}  & 0.1674 & 0.2643 & 0.2977 & 0.3247 & 0.3597 & 0.3548 & 0.4584 & 0.5655 & 0.7428 & 0.7232 \\ 
    & \textbf{TI-DiffRec (M)} & 0.1709 & 0.2757 & 0.3096 & 0.3378 & 0.3721 & 0.3723 & 0.4773 & 0.5887 & 0.7622 & 0.7407 \\ \cline{2-12} 
    & \textbf{SASRec (M)} & 0.2273 & 0.3532 & 0.3905 & 0.4190 & 0.4467 & 0.4674 & 0.5826 & 0.6955 & 0.8342 & 0.8007 \\ 
    & \textbf{+HBR}        & 0.2332 & 0.3597 & 0.3963 & 0.4250 & 0.4547 & \underline{0.4741} & 0.5872 & \underline{0.7006} & 0.8501 & \underline{0.8145} \\ 
    & \textbf{+HBR+NNS}     & \underline{0.2352} & \underline{0.3601} & \underline{0.3975} & \underline{0.4257} & \underline{0.4557} & 0.4733 & \underline{0.5890}  & 0.7002 & \underline{0.8517} & 0.8138 \\ 
    & \textbf{+HBR+NNS+DPA}  & \textbf{0.2363} & \textbf{0.3623} & \textbf{0.3992} & \textbf{0.4275} & \textbf{0.4572} & \textbf{0.4761} & \textbf{0.5904} & \textbf{0.7022} & \textbf{0.8520} & \textbf{0.8153} \\ \hline
\end{tabular}
\caption{Ablation versions of PDRec on two CDSR datasets. All improvements are significant compared to baselines.}
\label{tab:Ablation_CDSR}
\end{table*}

\subsection{Ablation Study (RQ2)}
\label{sec.ablation}
In this section, we conduct ablation studies to explore the effectiveness of different components in PDRec.
Thus we compare PDRec (SASRec) with different ablation versions of PDRec to verify the benefits of TI-DiffRec, HBR, DPN and NNS, respectively. Note that PDRec (SASRec) equals SASRec+HBR+NNS+DPA. From Fig. \ref{fig:ablation} we observe that:

(1) With HBR, SASRec+HBR achieves consistent improvement over SASRec. This mainly stems from the fact that diffusion-based preferences generated by the powerful TI-DiffRec can effectively denoise the historical behaviors via reweighting. It enables the recommender to emphasize the indispensable supervised signals while disregarding noisy interactions, thereby enhancing training efficiency.

(2) Comparing SASRec+HBR+NNS to SASRec+HBR, we find that NNS yields performance gains across most datasets. It demonstrates that these ``safe'' negative items judged by previous diffusion models could aid in alleviating the inherent false negative problems in model training.

(3) PDRec further improves the performance of SASRec+HBR+NNS. DPA emphasizes the top-ranked preferences determined by the diffusion model for unobserved items, thereby inferring user's more diverse potential preferences given by diffusion models. By double checking these high-quality positive augmentation candidates via self-distillation, DPA could bring in additional positive signals via a more flexible way to fight against data sparsity.

(4) PDRec achieves significant improvement compared to PDRec without TI-DiffRec (i.e., replacing TI-DiffRec with another SASRec). It highlights the necessity of employing diffusion models. Owing to the problem formulation, DM preserves the visibility into all items in the corpus. In conjunction with its powerful generalization ability, DM can offer informative knowledge relative to sequential models (i.e., SASRec), particularly for sparse user-item interaction matrices. Nevertheless, compared to the original DiffRec, PDRec is more effective and practical.

\subsection{Universality Analysis of PDRec (RQ3)}
\label{sec.uni_seq}
PDRec is a model-agnostic framework. To verify this, we employ each ablation variant of PDRec over GRU4Rec \cite{GRU4Rec} and CL4SRec \cite{CL4SRec} on Toy and Game datasets. Fig. \ref{fig:uni_seq} illustrates the results. We can find that:

(1) PDRec achieves significant improvements over different base models (GRU4Rec and CL4SRec) across diverse datasets. This demonstrates the universality of PDRec on different sequential encoders. Furthermore, it indirectly underscores the potential of PDRec to leverage the possible advancements in SR in the future, thereby extending the lifespan of the proposed DM-utilization frameworks.

(2) Progressive improvements are discernible among distinct versions of PDRec, with PDRec outperforming all its variants. It demenstrates that the proposed components are effective and universal for different base sequential encoders and datasets, further reconfirming the universality of PDRec.

\subsection{Results of Cross-domain SR (RQ4)}
\label{sec.uni_scen}

PDRec could also benefit positive transfer in CDSR. We follow typical CDSR settings \cite{Tri-CDR, DDGHM} and employ PDRec with SASRec (M) (M indicates directly mixing both source and target domains' behaviors in chronological order) on Toy$\rightarrow$Game and Game$\rightarrow$Toy settings. We also implement T-DiffRec (M) and TI-DiffRec (M) on the mixed (M) setting. From Table. \ref{tab:Ablation_CDSR}, we have:

(1) PDRec outperforms all diffusion-based models in CDSR, which implies that PDRec could be used in other tasks such as cross-domain scenarios. Its HBR provides an intuitive but effective way to filter negative transfers in cross-domain recommendation (i.e., mixing all domains' behaviors sequentially and conducting reweighting via diffusion), which could be further explored in the future.

(2) PDRec outperforms all of its ablation versions on most CDSR settings, with each component contributing to incremental improvements. It reconfirms the effectiveness and universality of HBR, NNS, and DPA from diffusion model.

(3) It is impressive that PDRec exhibits notable improvements across various metrics compared to the original T-DiffRec/TI-DiffRec on the mixed behavior sequence (\textbf{up to $\textbf{38.3\%}$}). It reiterates our main contribution that takes full advantage of the outputs of diffusion model as a plugin in SR.

\section{Conclusion}
In this paper, we propose an effective and model-agnostic Plug-in Diffusion Model for Recommendation (PDRec) framework. Instead of focusing on the highest-score item, PDRec fully leverages the diffusion-based preferences on all items. PDRec employs a historical behavior reweighting method to identify the indispensable behaviors and conducts a knowledge extracting strategy from both the unobserved items via the diffusion-based positive augmentation, and noise-free negative sampling. The extensive experiments and analyses on four datasets, three base models and two recommendation tasks demonstrate the effectiveness and universality of PDRec. In the future, we will continue to explore the tailored hard negative sampling strategies in PDRec and attempt to adapt PDRec as a flexible and detachable plugin in diverse recommendation scenarios.

\section{Acknowledgments}
This work is supported in part by the TaiShan Scholars Program (Grant no. tsqn202211289), the National Natural Science Foundation of China (Grant no. 62006141), the Excellent Youth Scholars Program of Shandong Province (Grant no. 2022HWYQ-048) and the Oversea Innovation Team Project of the  "20 Regulations for New Universities" funding program of Jinan (Grant no. 2021GXRC073). ChatGPT and Grammarly were utilized to improve grammar and correct spelling.

\bibliography{aaai24}
\end{document}